\documentclass[superscriptaddress,showpacs,preprint]{revtex4}
\usepackage[ansinew]{inputenc}
\usepackage [dvips]{graphicx}

\begin{document}
\title{Effect of annealing on carrier density and Curie temperature in epitaxial (Ga,Mn)As thin films}

\author{B. S. S\o rensen\footnote{E-mail: skov@fys.ku.dk}}
\affiliation{Niels Bohr Institute fAFG, \O rsted Laboratory,
University of Copenhagen, Universitetsparken 5. DK-2100
Copenhagen, Denmark.}

\author{J. Sadowski}
\affiliation{Niels Bohr Institute fAFG, \O rsted Laboratory,
University of Copenhagen, Universitetsparken 5. DK-2100
Copenhagen, Denmark.} \affiliation{MAX-Lab, Lund University,
SE-221 00 Lund, Sweden.} \affiliation{Institute of Physics, Polish
Academy of Sciences, al. Lotnik\`{o}w 32/46, PL-02-668 Warszawa,
Poland.}

\author{R. Mathieu\footnote{New adress: ERATO-SSS/JST, Cryogenic Center, University of Tokyo, 2-11-16 Yayoi, Bunkyo-ku, Tokyo 113-0032, Japan.}}
\affiliation{Department of Materials Science, Uppsala University, Box 534, SE-751, 21 Uppsala, Sweden.}

\author{P. Svedlindh}
\affiliation{Department of Materials Science, Uppsala University, Box 534, SE-751, 21 Uppsala, Sweden.}

\author{P. E. Lindelof}
\affiliation{Niels Bohr Institute fAFG, \O rsted Laboratory, University of Copenhagen, Universitetsparken 5.
DK-2100 Copenhagen, Denmark.}


\date{\today}

\begin{abstract}
We report a clear correspondence between changes in the Curie temperature and carrier density upon annealing in
epitaxially grown (Ga,Mn)As layers with thicknesses in the range between 5 nm and 20 nm. The changes are dependent
on the layer thickness, indicating that the (Ga,Mn)As - GaAs interface has importance for the physical properties
of the (Ga,Mn)As layer. The magnetoresistance shows additional features when compared to thick (Ga,Mn)As layers,
that are at present of unknown origin.
\end{abstract}

\maketitle The discovery of ferromagnetism in the dilute magnetic semiconductor (Ga,Mn)As~\cite{ohno} initiated a
rapid expansion in both the experimental and theoretical investigation of this material. Ferromagnetism in
(Ga,Mn)As is commonly related to the magnetic exchange interaction between the conduction holes and the magnetic
moment of the localized substitutional  Mn$^{+2}$ ions.~\cite{dietl} In the Zener model, the ferromagnetic ordering
temperature ($T_{c}$) is directly related to the content of Mn and the number of free charge carriers.~\cite{dietl}
The low temperature conditions necessary for epitaxial growth of (Ga,Mn)As, induces defects such as
Mn-interstitials~\cite{yu} and As-antisites~\cite{korzhavyi} that act as compensating donors and hence suppress
ferromagnetic ordering. Experimental work has shown that $T_{c}$, together with the electrical transport
properties, can be changed by post-growth annealing,~\cite{edmonds,edmondsA,ku,potashnik,kuryliszyn} and Curie
temperatures up to 150 K have now been reported.~\cite{ku} The main mechanism for this is believed to be a
reduction of Mn interstitials which are highly mobile in the annealing temperature range commonly used.~\cite{yu}
So far single (Ga,Mn)As layers with thicknesses in the nm range have not been investigated in great detail.
Matsukura \emph{et al.}~\cite{matsukuraA} reported a decrease in $T_{c}$ with increasing layer thickness, which is
also observed by Ku \emph{et al.},~\cite{ku} who also investigated the effect of annealing.

We have previously found that the effect of post-growth annealing on the shape of the magnetization curve and Curie
temperature, depends on the thickness of the (Ga,Mn)As layer.~\cite{mathieu} In this paper we present results of
the effect of post-growth annealing on the carrier density determined from Hall measurements in the same
Ga$_{0.95}$Mn$_{0.05}$As films. We found that the carrier density follows the changes in $T_{c}$ upon annealing, in
agreement with what is generally seen in (Ga,Mn)As.~\cite{potashnik,edmondsA,ku} Ferromagnetism has been observed
down to a thickness of 5 nm.~\cite{sorensen} In this sample, an unexplained feature in the magnetoresistance appear
at low temperatures. We show here that this feature is also present in thicker samples with thicknesses up to 20
nm.

The studied samples are epitaxially grown Ga$_{0.95}$Mn$_{0.05}$As layers with thicknesses in the range between 5
and 25 nm, embedded in low temperature (LT) GaAs.~\cite{sorensen,mathieu} For electrical transport measurements
mesas are defined in a 100 $\mu$m wide Hallbar geometry by UV-lithography and wet chemical etching with Au/Zn/Au
contacts. Annealing is performed at 240 $^{\circ}$C in a Nitrogen atmosphere. Two sets of samples were annealed for
one and two hours respectively, except the 15 nm thick layer that was annealed an additional hour without any
significant change in Curie temperature from that observed after two hours of annealing.~\cite{mathieu} The Hall
measurements are all standard 4-point dc-measurements performed at T=300 mK.

For (Ga,Mn)As a positive magnetoresistance in Hall measurements below $T_{c}$ at low fields, is evidence of
in-plane magnetic anisotropy.~\cite{matsukura} This is followed by a negative magnetoresistance that has been
understood  as a reduction of random spin-flip scattering with increasing magnetic field for metallic
samples.~\cite{matsukura,iye}

\indent The Hall resistance contains two contributions: ordinary and anomalous Hall coefficients and is given by:
\begin{equation}
R_{H}=\frac{R_{o}}{d}B+\frac{R_{S}}{d}M
\end{equation}
$R_{o}=1/pe$ is the ordinary Hall coefficient. $R_{S}$ is the anomalous Hall coefficient where Skew- and Side-jump
scattering contribute with terms given by $R_{S}\propto R_{sheet}$ and $R_{S}\propto R_{sheet}^2$ respectively. $M$
is the magnetization of the sample and $d$ is the thickness of the (Ga,Mn)As film. Recently a quadratic dependence
on the resistivity of the anomalous Hall effect consistent with a scattering independent anomalous Hall
conductivity has been reported.~\cite{edmonds}

Below $T_{c}$, the anomalous Hall term leads to a rapid increase in Hall resistance with increasing magnetic field,
often followed by a small decrease due to the negative magnetoresistance. Therefor it is often necessary to apply a
high magnetic field in order to determine the carrier density from the ordinary Hall-term.

\begin{figure}[e]
\includegraphics[width=.32\textwidth]{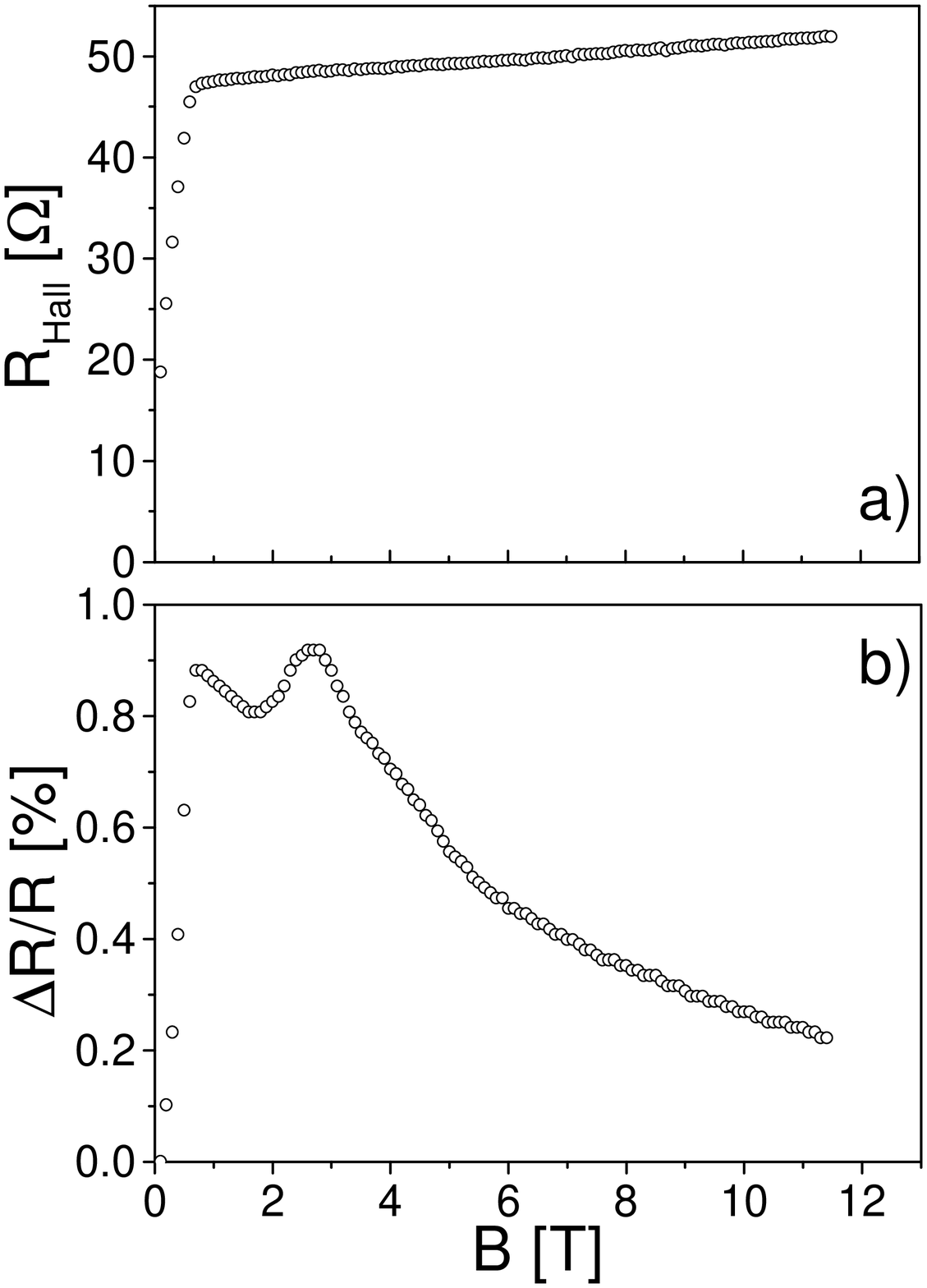}
\caption{\label{1}\ Hall- and magnetoresistance at T=300 mK of the 15 nm thick GaMnAs film annealed for 3 hours at
240 $^{\circ}$C.}
\end{figure}

\begin{figure}[e]
\includegraphics[width=.34\textwidth]{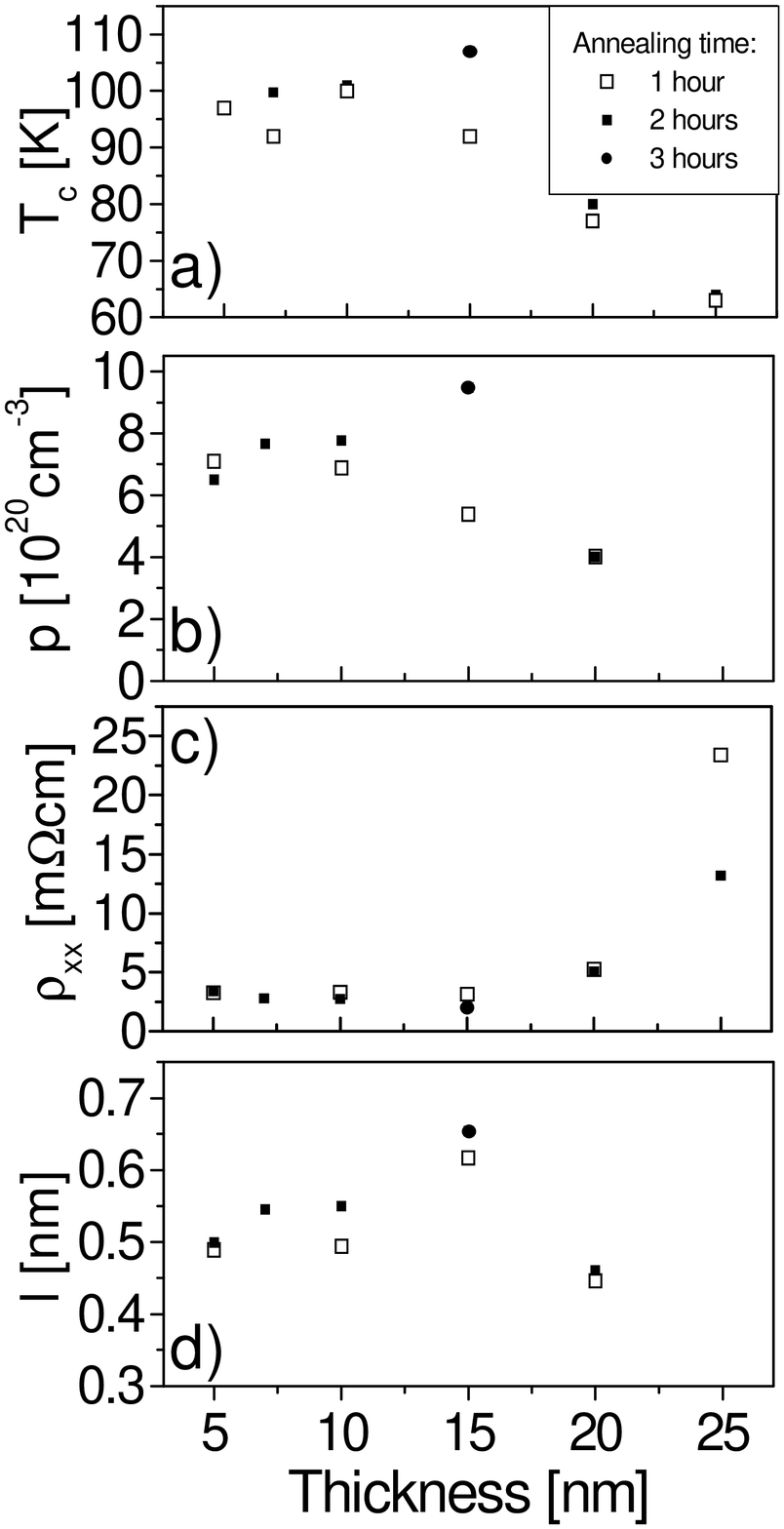}
\caption{\label{2}\ (Ga,Mn)As parameters at T=300 mK as a function of (Ga,Mn)As film thickness for samples annealed
at 240 $^{\circ}$C for one hour (open squares), 2 hours (filled squares) and 3 hours (circle). a) Curie
temperature; b) Carrier density as determined from Hall measurements assuming only Skew scattering; c) Resistivity;
d) Mean free path.}
\end{figure}

Figure 1 shows the Hall- and magnetoresistance of the 15 nm thick layer annealed for three hours. Beside the
effects described above, there is an additional oscillating feature at around 3 T and 4.5 T. This feature, varying
in magnitude, is seen in all the samples with thicknesses below 20 nm. Annealing has some effect on the magnitude
and magnetic field dependence of this oscillation, but no direct connection could be found. For some samples the
oscillation appears in the Hall resistance with the same magnetic field dependence as in the magnetoresistance,
probably originating from the anomalous Hall term. The oscillating feature is seen to disappear at higher
temperatures.

The carrier density (p) has been found by fitting the Hall data to eq. (1) for samples with (Ga,Mn)As layer
thickness up to 20 nm assuming either Skew or Side-jump scattering. The results assuming only Skew scattering are
shown in Fig. 2b. There is a small difference in the determined carrier density, which increases with increasing
layer thickness, depending on the assumed scattering mechanism where Skew scattering gives the highest values. The
negative magnetoresistance increases with increasing layer thickness, and the carrier density could not be
determined in the 25 nm thick layer, in the available magnetic field range. The carrier density increases upon
annealing except for the 5 nm layer, where it decreases. The increase upon annealing is consistent with the
reported decrease in the amount of Mn interstitials.~\cite{yu}

The Curie temperatures, shown in Fig. 2a, are also seen to increase during annealing. The Curie temperatures have
been determined by SQUID magnetometery~\cite{mathieu} except for the 5 nm thick layer, which was determined from
Hall measurements.~\cite{sorensen} This observation is contrary to recently reported results, where the presence of
a capping layer reduces T$_{c}$ in 10-50 nm thick (Ga,Mn)As layers upon annealing.~\cite{ku}

The mean free path (\emph{l}), calculated from the measured resistivity and carrier density at T=300 mK, is shown
in Fig. 2d. The obtained \emph{l} values are in the range between 0.4 nm and 0.65 nm, which is shorter than the
average distance between Mn-ions in (Ga,Mn)As with 5 $\%$ of Mn ($\sim$1.1 nm). These low values for the mean free
path is expected due to the high degree of disorder in low temperature grown (Ga,Mn)As. The increasing mean free
path upon annealing is a consequence of reduced disorder.

The clear correspondence between the carrier density and Curie temperature seen in Fig. 2a and Fig. 2b, suggests
that the Curie temperature is mainly governed by the number of carriers, and not directly related to the thickness
of the layer. In Fig. 3 the Curie temperature is plotted as a function of carrier density for the two sets of
samples, where the carrier density has been determined assuming either Skew- or Side-jump scattering. From a
log-log plot the Curie temperature is seen to follow a $p^{1/3}$ dependence to a good degree, when the carrier
density is determined by Skew scattering only. This dependence is less clear assuming only Side-jump scattering.
Having data only in the range of half a decade makes a fit to the relation: $T_{c}\propto p^{1/3}$, predicted by
Dietl \emph{et al.}~\cite{dietl} within the Zener model, questionable. Using a Mn content of 5$\%$ the term of
proportionality deviates by a factor of 2 from the value given by ref. ~\cite{dietl} when assuming Skew scattering.
This is acceptable considering the uncertainty on the Mn content and material parameters.

\begin{figure}
\includegraphics[width=.34\textwidth]{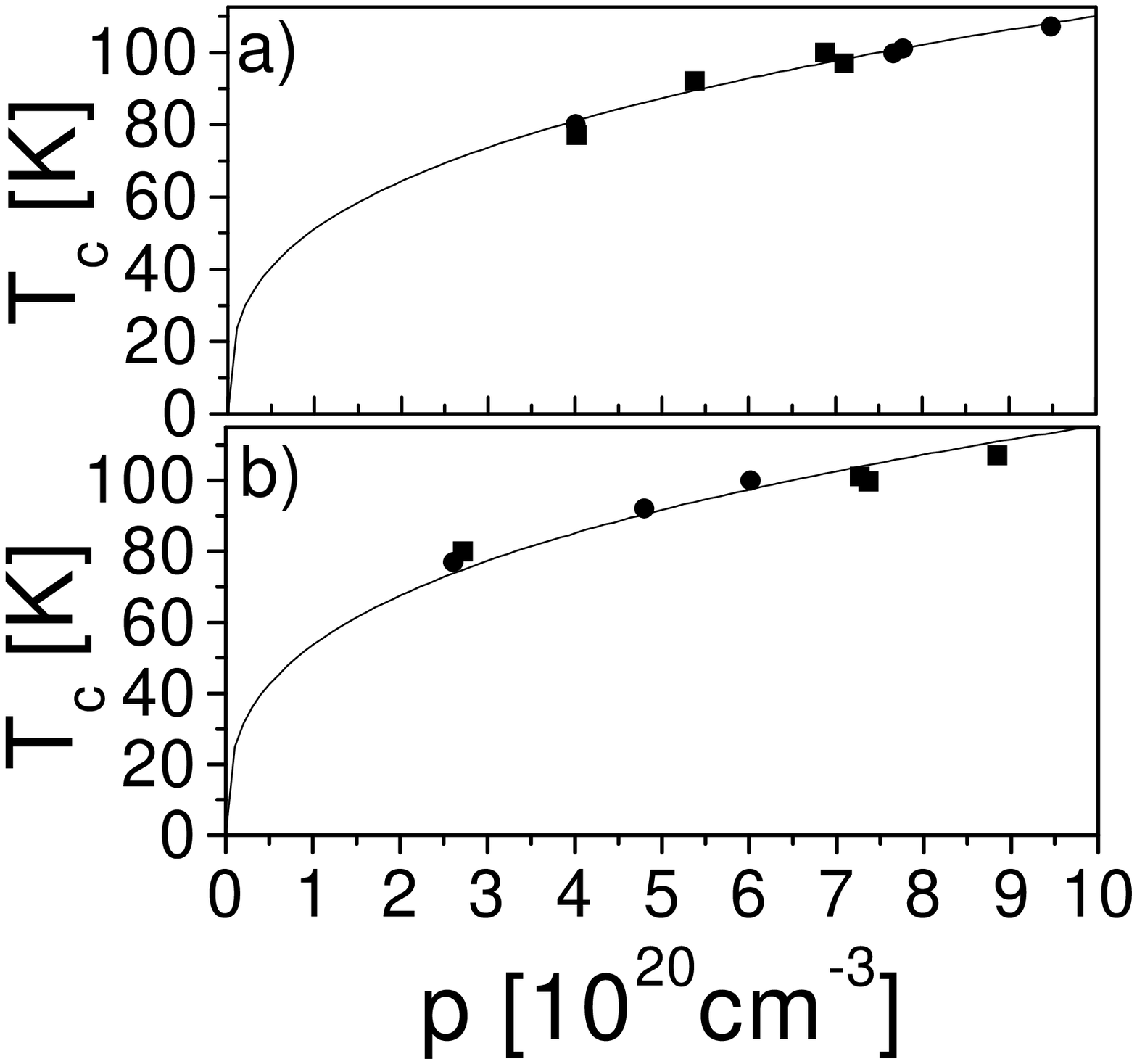}
\caption{\label{3}\ Curie temperature as a function of carrier density assuming only; a) Skew scattering and b)
Side-jump scattering, using the data from figure 2. The lines indicates a $T_{c}\propto p^{1/3}$ dependence.}
\end{figure}

Though it is difficult to speculate as to what mechanisms are important in (Ga,Mn)As thin films,~\cite{ku} these
findings suggest that the (Ga,Mn)As - LT GaAs interface plays an important role. Theoretical calculations show that
Mn ions in substitutional and interstitial positions have different energies near the surface,~\cite{erwin} and
experiments have found segregation of Mn on the surface of (Ga,Mn)As.~\cite{campion} This implies that annealing
may have a larger impact on (Ga,Mn)As layers with thickness comparable to the interface regions and that the Curie
temperature is mainly governed by the carrier density and hence the number of defects.

In summary, we have observed a clear correspondence between changes in Curie temperature and carrier density upon
annealing, suggesting that the impact of defects on the carrier density is dependent on the (Ga,Mn)As layer
thickness. A local maxima in the magnetoresistance at 3-4.5 T present only in layers with thickness below 20 nm is
observed, indicating that this is a thin film property.

Acknowledgement: This work was supported by the Swedish Natural Science Research Council (NFR), the Swedish
Research Council for Engineering Sciences (TFR), the Nanometer Structure Consortium in Lund, the Swedish Foundation
for Strategic Research (SSF), the Danish Research Council of Engineering Sciences (STVF framework programme
"Nanomagnetism"), and  the Danish Science Research Council (SNF framework programme "Mesoscopic Physics"). The
samples are partly prepared at the III-V NANOLAB.\newline

\end{document}